\documentclass[pre,preprintnumbers,amsmath,amssymb,showpacs,
nofootinbib,floatfix]{revtex4}

\usepackage{graphicx,bm}

\makeatletter
\def\graphicscale{\twocolumn@sw{0.33}{0.4}}
\def\graphicthreescale{\twocolumn@sw{0.33}{0.4}}

\begin{document}

\title{Invariant expectation values in the sampling of discrete frequency distributions}

 \author{Paolo Rossi}
   \affiliation{Dipartimento di Fisica dell'Universit\`a di Pisa and
   I.N.F.N., Sezione di Pisa, Largo Bruno Pontecorvo 3, I-56127 Pisa,
   Italy} \date{May 2, 2013}

\begin{abstract}
The general relationship between an arbitrary frequency distribution and the expectation value of the frequency distributions of its samples is discussed. A wide set of measurable quantities (``invariant moments") whose expectation value does not in general depend on the size of the sample is constructed and illustrated by applying the results to Ewens sampling formula. Invariant moments are especially useful in the sampling of systems characterized by the absence of an intrinsic scale. Distribution functions that may parametrize the samples of scale-free distributions are considered and their invariant expectation values are computed.  The conditions under which the scaling limit of such distributions may exist are described.

\end{abstract}


\maketitle


\section{Introduction}

In many interesting physical, biological and social phenomena, whenever no intrinsic scale for the relevant variables is present, the emergence of "scaling laws" is phenomenologically observed~\cite{Newman}.
However, strictly speaking, a power law is not a proper way of fitting empirical data, since no choice of the exponent can keep the higher moments of a power law distribution from diverging, while every phenomenological distribution leads to finite values for all moments.
This is not just a technicality: it is rather a reflection of the fact that the long tail of a power law distribution is in practice cutoffed by the existence of some "hidden" scale, irrelevant in the scaling region, but eventually forcing some upper limit on the variables describing the system.
It would therefore be convenient to parametrize the data by means of more regular distribution functions, sufficiently damped for very large values of the variables, but admitting power law distributions as regular limits when the control parameter implementing the cutoff is sent to its limiting value.

A related issue concerns the effects of sampling, which may be non trivial even when we restrict our attention to the expectation values of the sampled variables.
On average sampling does not affect the distributions of individual objects belonging to different kinds, but when we consider frequency distributions (that is the number of kinds that are represented $k$ times in a given population) we cannot in general expect that the frequency distribution in the samples be the same as in the original population, even after averaging on many different samples, basically because the cutoff induced by sampling acts differently (and in general nontrivially) at different scales.
It is therefore quite important  to be able to extract from the frequency distribution of the samples some information reflecting directly some intrinsic property of the underlying distribution.

Our purpose is therefore threefold. First we want to discuss the general relationship existing between some arbitrary frequency distribution and the expectation value of the frequency distributions of its samples, and construct observables whose expectation values turn out to be independent of the sample size, and therefore coinciding with the value taken by the same observables in the full distribution.

Moreover we want to study classes of distributions whose samples preserve the functional dependence on the parameters present in the original distribution, establishing the connection between the (a priori unknown) values of the parameters of the distribution and the (empirically measured) parameters of the sample distributions.

Finally we want to study the scaling limit of these distributions (when it exists), in order to explore the possibility of their use for the phenomenological description of systems that are theoretically expected to show scaling in the limit when all empirical cutoffs (including those induced by sampling) are going to disappear.

In Section \ref{framework} we establish the notation and the general framework of our analysis.

In Section \ref{moments} we construct a wide set of combinations  of expectation values that do not depend on the sample size.

In Section \ref{ewens} we apply our approach to the popular Ewens sampling formula, showing that its features  are consistent with the general pattern and computing its invariant expectation values.

In Section \ref{small} we consider the limiting case of a small sampling applied to a large population.

In Section \ref{large} we focus on the case when the original population and its samples are sufficiently large in comparison with typical frequency values, finding a useful mathematical relationship between the generating function of the expectation values of the sample distributions and the generating function of the original distribution.

In Section \ref{distributions} we analyze a class of  distributions (the so-called negative binomial distributions) admitting a scaling limit and enjoying the property that the distribution of expectation values of the samples has the same mathematical form as the original distribution. We also compute in a closed form the values ot the basic invariant expectation values for these distributions.

Finally in Section \ref{scaling} we analyze  the scaling limit itself and discuss the conditions under which one may expect this limit to be a sensible description of the original system.

Appendices are devoted to the proofs of some mathematical results and to discussing the issue of correlation between random samples.

\section{The general framework}
\label{framework}
We are considering a set of $N$ objects (``individuals'') belonging to $S$ different kinds (``species''), and we assume that the set contains $\hat N_a$ objects of the $a$-th kind, subject to the constraint $\sum_a \hat N_a = N$.

A sample is a set of $n$ objects, containing $\hat n_a$ objects of the $a$-th kind, subject to the constraint $\sum_a \hat n_a = n$.

The probability $P_{\{\hat n_a\}}$ of extracting a specific sample $\{\hat n_a\}$ from a given set $\{\hat N_a\}$ is obtained from the multivariate hypergeometric distribution
$$P_{\{\hat n_a\}} = {N \choose n}^{-1}\prod_{a=1}^S {\hat N_a \choose \hat n_a}.$$

We can easily compute the relevant expectation values, obtaining in particular
$$\langle \hat n_a\rangle  = \hat N_a {n \over N}  = n\,\hat p_a, \qquad \qquad \langle \hat n_a^2\rangle -\langle \hat n_a\rangle ^2 = {N-n \over N-1} n \hat p_a (1-\hat p_a)$$
where $\hat p_a \equiv {\bar N_a/N}$ is the probability of extracting an object of the $a$-th kind in a single extraction.

It may be useful to consider also the limit of small samples $\hat n_a \ll  \hat N_a$. In this limit the probability of a specific sample is well approximated by the multinomial distribution
$$P_{\{\hat n_a\}} = n! \prod_{a=1}^S {1 \over \hat n_a!} (\hat p_a)^{\hat n_a}.$$

A frequency distribution is a set of values $\{N_k\}$, where $N_k$ is the number of kinds such that for each of them there are $k$ objects in the original set. According to the definition, the following conditions must be satisfied:
$$\sum_{k=1}^N N_k = S, \qquad \qquad \qquad \sum_{k=1}^N k\,N_k = N.$$

The frequency distribution of a sample is a set of values $\{n_l\}$, satisfying the conditions
$$\sum_{l=0}^n n_l = S, \qquad \qquad \qquad \sum_{l=1}^n l\,n_l = n.$$

Notice that the frequency distribution of a sample formally includes the (unobservable) value $n_0$,  corresponding to the number of kinds, present in the original set, which are not represented in the sample.

It is in principle possible to compute the probability of any sample distribution $\{n_l\}$ as a function of a given set $\{N_k\}$. To this purpose it is convenient to define the intermediate variables $N_{kl}$, representing the (random) number of kinds characterized by $k$ objects in the original set and by $l$ ($l \leq k$) objects in the sample. The variables $N_{kl}$ are strongly constrained, since they must satisfy all the conditions: 
$$\sum_{l=0}^n N_{kl} = N_k, \qquad \qquad \qquad \sum_{k=1}^N N_{kl} = n_l.$$
\smallskip

The probability $P_{\{N_{kl}\}}$ of a specific configuration ${\{N_{kl}\}}$ follows from the general probability formula~\cite{Zelterman}:
$$P_{\{N_{kl}\}} ={N \choose n}^{-1} \prod_{k=1}^N \biggl[N_k! \prod_{l=0}^k {1 \over N_{kl}!} {k \choose l}^{N_{kl}} \biggr],$$
subject to the constraint $\sum_{l=0}^n N_{kl} = N_k$.

The probability $P_{\{n_l\}}$ of finding a frequency distribution $\{n_l\}$ in a sample is obtained by summing the probabilities $P_{\{N_{kl}\}}$ over all configurations  satisfiying the constraint $\sum_{k=1}^N N_{kl} = n_l$. The corresponding multivariate generating function can be defined as
$$\varepsilon^{(n)}(\{t_l\}) \equiv \sum_{\{n_l\}} P_{\{n_l\}} \prod_{l=0}^n  t_l^{n_l} =\sum_{\{N_{kl}\}} P_{\{N_{kl}\}} \prod_{k=1}^N \prod_{l=0}^k  t_l^{N_{kl}}.$$

It is also possible (and it will be quite convenient) to define a cumulative generating function for the probability of finding the frequency distributions $P_{\{n_l\}}$ for samples of all possible sizes :
$$E(x;\{t_l\}) \equiv  \sum_{n=0}^N  {N \choose n}\varepsilon^{(n)}(\{t_l\}) x^n =  \sum_{\{N_{kl}\}} \prod_{k=1}^N \Bigl(N_k! \prod_{l=0}^k   {1 \over N_{kl}!}\Bigl[ {k \choose l} x^l t_l \Bigr]^{N_{kl}}  \Bigr) = \prod_{k=1}^N \Bigl[ \sum_{l=0}^k  {k \choose l}x^l t_l \Bigr]^{N_k},$$
where we used the explicit expression of $P_{\{N_{kl}\}}$ and all the relevant constraints.

The expectation values $\langle n_l\rangle $ can be computed starting from the above expressions and from the relationship
$$\langle n_l\rangle  = \sum _{k=1}^N \langle N_{kl}\rangle  =\sum_{k=1}^N \sum_{\{N_{jm}\}}  N_{kl} P_{\{N_{jm}\}}.$$

Straightforward manipulations lead to the results~\cite{Zelterman}
$$\langle N_{kl}\rangle  = N_k {{k \choose l}{N-k \choose n-l} \over {N \choose n}}, \qquad \qquad \langle n_l\rangle  = {\sum_{k=1}^N N_k {k \choose l}{N-k \choose n-l}  \over {N \choose n}}.$$

It is easy to check that the following relationships are satisfied:
$$\sum_{l=0}^n \langle n_l\rangle  = \sum_{k=1}^N N_k = S, \qquad \qquad
\sum_{l=0}^n l\, \langle n_l\rangle  = \Bigl(\sum_{k=1}^N k\, N_k \Bigr){n \over N} = n.$$

In order to fully appreciate the relevance of considerations based on the expectation values we must evaluate the weight of the fluctuations.  Taking second derivatives  of the generating function $E(x; t_l)$ one obtains:
$$\langle n_l^2\rangle -\langle n_l\rangle ^2 = \sum_{k,k'} N_k N_{k'}  {k \choose l} {k' \choose l} \Biggl[{ {N-k-k' \choose n-2 l} \over {N \choose n}}-{ {N-k \choose n-l} \over {N \choose n}}{ {N-k' \choose n-l} \over {N \choose n}} \Biggr] +\sum_k N_k  {k \choose l}  \Biggl[{ {N-k \choose n- l} \over {N \choose n}}- {k \choose l} { {N-2 k \choose n-2 l} \over {N \choose n}} \Biggr].$$

Notice  that in the large $N$ limit the term quadratic in $N_k$ is depressed by a power of $1/N$. This observation suggests that  very important limits of the above results may be obtained when considering  large populations.

\section{Invariant expectation values}
\label{moments}

It is very important to be able to define a set of expectation values that are independent of the size of the sample, and therefore may reflect very directly the properties of the original frequency distribution.

Let's consider the following combinations of expectation values:
$$\langle m_{\{p_i\}}^{(n)}\rangle  \equiv {n \choose P}^{-1}\sum_{\{q_i\}} \prod_{i=1}^I \Bigl[{q_i \choose p_i} {\partial \over \partial t_{q_i}}\Bigr]  \varepsilon^{(n)}(\{t_l\}|_{\{t_l=1\}},$$
where $p_i$ are $I$ arbitrary positive integer numbers, subject only to the constraint that $P \equiv \sum_i p_i \leq n$.

The definition of the quantities appearing in the r.h.s. implies that the derivatives with respect to $t_{q_i}$ are the joint factorial moments of the distribution; therefore we are dealing with weighted combinations of joint factorial moments. When some of the indices $p_i$ are equal to one, the expectation values may be expressed in terms of a combination of lower rank moments ($I'\langle I$).

It is possible to recognize that the quantities $\langle m_{\{p_i\}}^{(P)}\rangle $ are related (up to a trivial combinatorial factor taking into account the existence of $n_p$ coincident values of the indices $p_i$) to the probability of finding the configurations $\{p_i\}$ in the sample containing $P$ elements, and are therefore strictly connected with the probabilities $P_{\{n_p\}}$ .

Exploiting the properties of the generating function $E(x; \{t_l\})$ we prove in Appendix \ref{Expectation} that
$$\langle m_{\{p_i\}}^{(n)}\rangle  =\langle  m_{\{p_i\}}^{(N)}\rangle  \equiv M_{\{p_i\}}$$
for all sets $\{p_i\}$ such that $P \leq n$. Hence the expectation values of the nontrivial  invariant moments $m_{\{p_i\}}$ evaluated for samples of arbitrary size $n \geq P$, coincide with the moments $M_{\{p_i\}}$ of the original frequency distribution. If the original set was generated by a random process, also the  $M_{\{p_i\}}$ will be expectation values.
Recalling that $\varepsilon^{(N)}(\{t_l\}\equiv \prod_k t_k^{N_k}$ we may now generate a representation of all $P_{\{n_p\}}$ in terms of $N_k$, without making use of the coefficients $N_{kl}$.

The properties of the binomial coefficients make it possible to invert the relationship between invariant moments and joint factorial moments, thus finding that
$$\Bigl[\prod_{i=1}^I {\partial \over \partial t_{q_i}}\Bigr]  \varepsilon^{(n)}(\{t_l\}|_{\{t_l=1\}} = \sum_{\{p_i\}}\prod_{i=1}^I (-1)^{p_i-q_i}{p_i \choose q_i}\langle {n \choose P} m_{\{p_i\}}^{(n)}\rangle  = \sum_{\{p_i\}}\prod_{i=1}^I (-1)^{p_i-q_i}{p_i \choose q_i}{n \choose P} M_{\{p_i\}}.$$

The basic invariant moments are
$$m_p^{(n)} = {n \choose p}^{-1} \sum_{q=p}^n  { q \choose p}n_q.$$

According to the inversion formula 
$$\langle n_l\rangle  = {\partial \varepsilon^{(n)} \over \partial t_l}|_{\{t_m =1\}} = \sum_{p=l}^n (-1)^{p-l}{p \choose l}{n \choose p}\langle m_p^{(n)}\rangle  =\sum_{p=l}^n (-1)^{p-l}{p \choose l}{n \choose p} M_p.$$

One may define generating functions for the expectation values of $n_l$ and of the basic invariant moments:
$$f^{(n)}(t) \equiv \sum_{l=0}^n \langle n_l\rangle  t^l, \qquad \qquad g^{(n)}(z) \equiv f^{(n)} \bigl(1+{z \over n}\bigr) = \sum_{p=0}^n {n \choose p} M_p \Bigl({z \over n}\Bigr)^p.$$

Notice that a special case of the above formula is
$$F(t) \equiv \sum_{k=0}^N N_k t^k, \qquad \qquad G(z) \equiv F\bigl(1+{z \over N}\bigr)= \sum_{p=0}^N {N \choose p} M_p \Bigl({z \over N}\Bigr)^p.$$

It is immediate to recognize that $g^{(n)}(0) =G(0)= M_0 \equiv S$, and ${dg \over dz}^{(n)}(0) = {dG \over dz}(0) = M_1 \equiv 1$.

\section{Application to Ewens sampling  formula}
\label{ewens}

The multivariate Ewens distribution~\cite{Ewens,Karlin}, called in genetics the Ewens sampling formula, describes a specific probability for the partition of $n$ into parts, and found its main applications in the context of the neutral theory of evolution and in the unified neutral theory of biodiversity~\cite{Hubbell,Rosindell}. Since Ewens formula and its possibile generalizations have been the subject of a wide and  still growing literature~\cite{Griffiths, Etienne, Lessard, Lambert}, it may be interesting to apply the results presented in Section \ref{moments} to this specific instance. In our notation Ewens probability distribution takes the form
$$P_{\{n_l\}} ={1 \over \aleph_n} \prod_{l=1}^n {1 \over n_l!}\Bigl({\theta \over l}\Bigr)^{n_l},\qquad \qquad \qquad \aleph_n \equiv {\Gamma(\theta+n) \over n! \,\Gamma(\theta)},$$
where $0< \theta < \infty$ and $\sum l\,n_l = n$.

Joint factorial moments of the Ewens distribution are easily computed~\cite{Johnson} and one can show that
$$\prod_{i=1}^I \Bigl[ {\partial \over \partial t_{q_i}}\Bigr] \sum_{\{n_l\}} P_{\{n_l\}} \prod_{l=0}^n  t_l^{n_l}= {\aleph_{n-Q} \over \aleph_n}\prod_i \Bigl({\theta \over q_i}\Bigr) ,$$
where $Q \equiv \sum_i q_i \leq n$.

We are then left with the task of computing the summations appearing in the equation
$$\langle m_{\{p_i\}}^{(n)}\rangle   ={n \choose P}^{-1} \sum_{\{q_i \}} \prod_{i=1}^I { q_i \choose p_i}{\aleph_{n-Q} \over \aleph_n}\prod_i \Bigl({\theta \over q_i}\Bigr) = {P! \,(n-P)!\,\Gamma(\theta)  \over \Gamma(\theta + n)} \prod_{i=1}^I \Bigl({\theta \over p_i}\Bigr) \sum_{\{q_i \}} \prod_{i=1}^I { q_i-1 \choose p_i-1} {\Gamma(\theta+n-Q) \over \Gamma(\theta)\,(n-Q)!}.$$

We prove in Appendix \ref{Combinatorics} that
$$\sum_{\{q_i \geq p_i \}} \prod_{i=1}^I { q_i-1 \choose p_i-1} {\Gamma(\theta+n-Q) \over \Gamma(\theta)\,(n-Q)!} = {\Gamma(\theta+n) \over \Gamma(\theta+P)\,(n-P)!},$$
hence
$$\langle m_{\{p_i\}}^{(n)}\rangle  =  {P! \,\Gamma(\theta) \over \Gamma(\theta+P)}\prod_{i=1}^I \Bigl({\theta \over p_i}\Bigr) \equiv {1 \over \aleph_P}\prod_{i=1}^I \Bigl({\theta \over p_i}\Bigr),$$
showing explicitly that the expectation values of the invariant moments of the Ewens distribution are independent of the sample size and related to the probability of the configuration $\{p_i\}$ in the sampling of $P$ elements.

We stress that invariant moments, because of their independence from the size of the sample, may become a highly valuable tool in testing the applicability of Ewens distribution (and of the conceptual assumptions underlying its derivation) to the interpretation of actual empirical data.

\section{Large population and small samples}
\label{small}

A significant simplification occurs when $N \rightarrow \infty$ while all other variables are kept finite. 
Setting $\tilde x \equiv Nx$ and $t_0 = 1$ in the cumulative generating function and taking the large $N$ limit we obtain
$$\tilde E(\tilde x;\{t_l\}) \equiv 1+\sum_{n=1}^\infty \tilde \varepsilon^{(n)} (\{t_l\}) {\tilde x^n \over n!} =\prod_{k=1}^\infty \Bigl[1+\sum_{l=1}^k {k \choose l} \Bigl({\tilde x \over N}\Bigl)^l t_l\Bigr]^{N_k} \rightarrow \prod_{k=1}^\infty \Bigl[1+\sum_{l=1}^\infty \Bigl( {k\,\tilde x \over N}\Bigr)^l {t_l \over l!}\Bigr]^{N_k}.$$

Let's now define (for $n,l$ different from zero) the following set of coefficients:
$$c^{(n)}(\{n_l\}) \equiv (-1)^{s-1} (s-1)!\prod_{l=1}^n {1 \over n_l!}\Bigl({1 \over l!}\Bigr)^{n_l},$$
where $s= \sum_l n_l$ and $n = \sum_l l\,n_l$, and notice that the definition of $\tilde E$ implies that
$$\ln \tilde E(\tilde x;\{t_l\})  \equiv \sum_{n=1}^\infty \Bigl[\sum_{\{n_l\}} c^{(n)}(\{n_l\}) \prod_{l=1}^n \Bigl(\tilde \varepsilon^{(l)}(\{t_l\}) \Bigr)^{n_l}\Bigr] \tilde x^n.$$ 

It is also possible to recognize that, under the same assumptions,
$$\ln \tilde E(\tilde x;\{t_l\}) = \sum_{n=1}^\infty \Bigl[\sum_{\{n_l\}} c^{(n)}(\{n_l\}) \prod_{l=1}^n t_l^{n_l} \Bigr] \tilde m_n^{(N)} x^n,$$
where we introduced  the large $N$ limit of the basic invariant moments: $\tilde m_p^{(N)} \equiv \sum_q N_q (q/ N)^p$.

Comparing the two results we conclude that, for each value of $n \rangle  0$,
$$\sum_{\{n_l\}} c^{(n)}(\{n_l\}) \prod_{l=1}^n \Bigl(\tilde \varepsilon^{(l)}(\{t_l\}) \Bigr)^{n_l}=\Bigl[\sum_{\{n_l\}} c^{(n)}(\{n_l\}) \prod_{l=1}^n t_l^{n_l} \Bigr] \tilde m_n^{(N)}.$$

These equations allow in principle for the recursive determination of all $\tilde \varepsilon^{(n)}(\{t_l\}$ in terms of $\{\tilde m_p^{(N)}\}$ (with $p\leq n$), starting from the initial condition $\tilde \varepsilon^{(1)} = t_1$. Higher rank invariant moments ($I \rangle  1$)  in the large $N$ limit become polynomials in the basic moments. However one must keep in mind that, when the set $\{N_k\}$ not fixed, but generated by a probability distribution (as in the case of the Ewens formula), the expectation values of the products of basic moments appearing in the l.h.s. do not coincide with the products of the expectation values.

\section{Large populations and large samples}
\label{large}

When $k,l \ll  N,n$  one may systematically exploit the property that, for small $a$ and $b$, 
$$ {N-a  \choose n-b } \rightarrow { \rho^b (1-\rho)^{b-a} \over \rho^n (1-\rho)^{N-n}} \qquad \qquad \qquad \rho \equiv {n \over N}.$$

Expressing $N$ and $n$ in terms of $N_{kl}$ one may then obtain
$$P_{\{N_{kl}\}}  \rightarrow \prod_{k=1}^N \biggl[N_k! \prod_{l=0}^k  {1 \over N_{kl}!} P_{kl}^{N_{kl}}\biggr] \qquad \qquad \qquad P_{kl} \equiv  {k \choose l} \rho^l (1-\rho)^{k-l} .$$

As shown in Appendix \ref{Fluctuations}, when computing expectation values with the above probability distribution, the constraint $\sum l\,n_l = n$ becomes irrelevant in the  large $N$ limit, and expectation values of products of $N_{kl}$ with different values of the index $k$ factorize into products of expectation values computed for each separate value of $k$. 
We can therefore compute directly the generating function for a fixed sample size, generalizing the multivariate multinomial distribution:
$$\varepsilon^{(n)}(\{t_l\}) = \prod_{k=1}^N \sum_{\{N_{kl}\}} \biggl[N_k! \prod_{l=0}^k {1 \over N_{kl}!}(P_{kl} \,t_l)^{N_{kl}}\biggr] =\prod_{k=1}^N \Bigl[ \sum_{l=0}^k P_{kl} \,t_l\Bigr]^{N_k}.$$

A consistency check is easily obtained by observing that $\varepsilon^{(n)}({1}) = 1$, because of the property that $\sum_{l=0}^k P_{kl} = 1$. 

The expectation value of $n_l$  turns out to be:
$$\langle n_l\rangle  =  \sum_{k=1}^N N_k  P_{kl},$$
and one may check that the conditions on $\sum_l \langle n_l\rangle $ and on $\sum_l l\, \langle n_l\rangle $ are still satisfied.

We can also estimate the behavior of  fluctuations when $k,l \ll  N,n$:
$$\langle n_l^2\rangle  - \langle n_l\rangle ^2 = \sum_{k=1}^N  N_k P_{kl}(1 -P_{kl}).$$

The above expression is always smaller than $\langle n_l\rangle $ and as a consequence fluctuations become unimportant for sufficiently large values of $\langle n_l\rangle $.

In the same limit we may derive a notable relationship between the generating function of the original frequency distribution and the generating function of the expectation values of its samples. In fact we may recognize that for sufficiently large $N$ and $n$
$$g^{(n)}(z)  = \sum_{p=0}^\infty  M_p {z^p \over p!}= G(z).$$

Since in general $f^{(n)}(t) = g^{(n)}\bigl(n(t-1)\bigr)$ and $F(t) = G\bigl(N(t-1)\bigr)$, it is then easy to check that
in the limit under consideration 
$$f^{(n)}(t) = F(1-\rho+\rho\,t),$$

As a direct consequence of these results, whenever the (size-independent) function $\gamma(z) \equiv G(z)-G(0)$ can be cast into a form exhibiting no explicit parametric dependence on $N$, the expectation values  $\langle n_l\rangle $  can be obtained from $N_k$ by the replacement $N \rightarrow n$.  

Notice that in the limit $k,l \ll  N,n$ the definition of the basic invariant moments $m_p^{(n)}$ simplifies to
$$m_p^{(n)} \rightarrow {p! \over n^p}\sum_{l=p}^n n_l {l \choose p}.$$

It is worth analyzing in this limit the explicit expressions of the second basic invariant moment:
$$M_2 \rightarrow {1 \over N^2} \sum_{k=1}^N  k(k-1) N_k = \sum_{a=1}^S \bigl({\hat N_a \over N}\bigr)^2 -{1 \over N} = \sum_{a=1}^S \langle \bigl({\hat n_a \over n}\bigr)^2\rangle  -{1 \over n} \equiv {1 \over \alpha}.$$

As shown in Appendix \ref{Correlation} the above results may be used also in order to parametrize the expected correlation between samples under the assumption of independent random sampling.

\section{A class of distributions and its properties}
\label{distributions}

As mentioned in the Introduction, distributions found in samples  may often correspond to systems whose asymptotic ($N \rightarrow \infty$)  distribution is expected to obey a scaling law. However the exponent of the scaling law will in general be nontrivial, in contrast with the prediction offered by the simplest neutral models. An example of empirical and theoretical evidence for nontriviality is offered by surname frequency distributions (see Ref.~\cite{Rossi} for a recent review),  recalling that surnames are expected to mimick the behavior of selectively  neutral alleles. it is therefore especially interesting to consider parametrizations that may reflect notriviality of exponents, ad  in particular the class of negative binomial distributions~\cite{Hilbe}, which can be obtained starting from the generating function

$$ F_c (t) = {N \over x}{(1-x)^{1-c} \over c} \bigl[1-(1-x t)^c \bigr] = \sum_{k=1}^\infty {N \over x}{(1-x)^{1-c} \over \Gamma(1-c)}{\Gamma(k-c) \over k!} (xt)^k,$$
where $0\langle x\langle 1$ and the parameter c is assumed to vary in the range $0 \leq c \langle 1$.

The asymptotic behaviour of the distribution for large $k$ is easily obtained by observing that in this limit
$${\Gamma (k-c) \over k!} \rightarrow {1 \over k^{1+c}}, \qquad \qquad
N_k \rightarrow {N \over x}{(1-x)^{1-c} \over \Gamma(1-c)}{x^k \over k^{1+c}}.$$

We can now compute the generating function for the expectation values of the samples according to the general rule previously discussed, and obtain
$$f_c(t) =f_c(0)+{n \over y}{(1-y)^{1-c} \over c} \bigl[1-(1-y t)^c \bigr],$$
where we have defined $y = {\rho x \over 1-x+\rho x}$.

The distribution of the samples has the same form as the original distribution, once the replacements $N \rightarrow n$ and $x \rightarrow y$ have been performed, and therefore we obtain the asymptotic behaviour
$$n_l \rightarrow {n \over y}{(1-y)^{1-c} \over \Gamma(1-c)}{y^k \over k^{1+c}}.$$

It is possible to define a combination of parameters independent of the dimension of the sample:
$$\beta = N{1-x \over x} = n{1-y \over y}.$$

It is useful to represent $x$ and $y$ in a form showing explicitly their dependence on the dimension of the sample and on the invariant parameter $\beta$:
$$x = {N \over \beta+N}, \qquad \qquad \qquad y = {n \over \beta+n}.$$

It is now possible to evaluate the expectation value of the  invariant moments from the expression
$$\gamma_c(z) \equiv G_c(z)-G_c(0) = {\beta \over c}\Bigl[ 1-\Bigl(1+{ z \over \beta}\Bigr)^c\Bigr],$$
showing no explicit parametric dependence on $N$; we therefore obtain (for $p \neq 0$)
$$M_p = \beta^{1-p} {\Gamma(p-c) \over \Gamma(1-c)}  \rightarrow {\beta^{1-p} \over \Gamma(1-c)} {p! \over  p^{1+c}}, \qquad \qquad \alpha \equiv {1 \over M_2} = {\beta \over 1-c}.$$

The limit of the above results when $c \rightarrow 0$ is smooth, and it corresponds to Fisher distribution~\cite{Fisher}, such that
$$F_0(t) = - \beta \ln (1-x t), \qquad \qquad \qquad N_k = \beta {x^k \over k},$$
and
$$f_0(t) = \beta  \ln \bigl({\beta+N \over \beta+ n}\bigr) -\beta \ln (1-y t), \qquad \qquad n_l = \beta {y^l \over l}.$$

The generating function of the invariant moments is obtained from $\gamma_0(z) =  - \beta \ln (1+ z/\beta)$,
and as a consequence the expected values of the invariant moments ($p \neq 0$) are exactly $M_p =(p-1)!\, \beta^{1-p}$.

Notice in particular the relationship $\beta = \alpha$, peculiar to Fisher distribution. By comparing these results with the large $\theta$ limit of the invariant moments of Ewens distribution we may easily check Watterson's~\cite{Watterson} and Hubbell's~\cite{Hubbell} observation that in this limit $\theta$ is strictly connected to Fisher's $\alpha$.

\section{The scaling limit}
\label{scaling}

Let's now consider very large systems, and assume that we can gather information only through the sampling of
$n$ objects belonging to the system, with $n$ large but not necessarily comparable to $N$.

The analysis of the invariant moments may then allow us to check the applicability of a phenomenological description of the samples based on some distribution falling into the classes discussed in the previous Sections.
In the case of a positive response to the check it is then possible to find numerical estimates of the parameter $\beta$ and of the exponent $c$.
Such estimates are clearly meaningful only if $\beta$ does not turn out to be significantly greater than $n$.

Under these assumptions, we can infer a description of the original system, and in the case $N \rangle \rangle  n$ such a description will correspond to computing the limit $x \rightarrow 1$ of the previous results.
As a consequence, at least for observable (i.e. not too large) values of $k$, the original distribution is expected to be well described by the scaling form
$$N_k \rightarrow  {N^c \beta^{1-c} \over \Gamma(1-c)}{1 \over k^{1+c}}.$$

\appendix

\section{Expectation values of the invariant moments}
\label{Expectation}

Let's consider the cumulative generating function for the expectation value of a given invariant moment (setting $P =\sum p_i$) for samples of all possible sizes:
$$\sum_{n=0}^N {N \choose n}{n \choose P}\langle m_{\{p_i\}}^{(n)}\rangle  x^n \equiv \sum_{\{q_i\}} \prod_{i=1}^I \Bigl[{q_i \choose p_i}{\partial \over \partial t_{q_i}}\Bigr]E(x;\{t_l\})|_{\{t_l=1\}}.$$

Let's now observe that, due to the property that
$$\ln E(x;\{t_l\}) = \sum_{k=1}^N N_k \ln  \Bigl[ \sum_{l=0}^k  {k \choose l}t_lx^l \Bigr],$$
the derivatives appearing in the above defined cumulative generating function, computed at ${t_l=1}$, can be expressed as summations (over $\{k_i\}$ indices) of products of $N_{k_i}$ times a universal  $x$-dependent factor
$$ \Bigl[\prod_{i=1}^I {k_i \choose q_i}\Bigr] x^Q (1+x)^{N-K},$$
where $Q = \sum_i q_i$ and  $K = \sum_i k_i$. However the following identity holds:
$${q_i \choose p_i}{k_i \choose q_i} = {k_i \choose p_i}{k_i-p_i \choose q_i-p_i}.$$

As a consequence the cumulative generating function is proportional to the factor
$$ \Bigl[\prod_{i=1}^I {k_i \choose p_i}\Bigr] x^P (1+x)^{N-K} \sum_{\{q_i\}}\Bigl[\prod_{i=1}^I {k_i-p_i \choose q_i-p_i} x^{Q-P}\Bigr]= \Bigl[\prod_{i=1}^I {k_i \choose p_i}\Bigr] x^P (1+x)^{N-P}.$$

The summations over the indices $\{k_i\}$ may now be formally performed, and, by matching the coefficient of $x^N$ in the two sides of the equation, the result can be easily recognized to coincide with the expected value of the invariant moment computed for the original distribution times the combinatorial factor ${N \choose P}$.

In conclusion we find
$$\sum_{n=0}^N {N \choose n}{n \choose P}\langle m_{\{p_i\}}^{(n)}\rangle  x^n =  {N \choose P}\langle  m_{\{p_i\}}^{(N)}\rangle  x^P (1+x)^{N-P}.$$

Expanding the r.h.s. in powers of $x$ and noticing that ${N \choose n}{n \choose P}= {N \choose P}{N-P \choose n-P}$
we finally obtain
$$\sum_{n=0}^N {N \choose P}{N-P \choose n-P}\langle m_{\{p_i\}}^{(n)}\rangle  x^n = \sum_{n=0}^N {N \choose P} {N-P \choose n-P} \langle m_{\{p_i\}}^{(N)}\rangle  x^n,$$
implying immediately that, as long as $n \geq P$,
$$\langle m_{\{p_i\}}^{(n)}\rangle  =\langle m_{\{p_i\}}^{(N)}\rangle  \equiv   M_{\{p_i\}}.$$

\section{Proof of a combinatorial formula}
\label{Combinatorics}

The identity
$$\sum_{\{q_i \geq p_i \}} \prod_{i=1}^I { q_i-1 \choose p_i-1} {\Gamma(\theta+n-Q) \over \Gamma(\theta)\,(n-Q)!} = {\Gamma(\theta+n) \over \Gamma(\theta+P)\,(n-P)!},$$
where $Q= \sum_i q_i$ and $P = \sum_i P_i$, can be proven by recalling that for real numbers $\alpha$ and positive integers $q \geq p$
$$(1-x)^{-\alpha} = \sum_{m=0}^\infty {\Gamma(\alpha + m) \over \Gamma(\alpha)\,m!}x^m, \qquad \qquad (1-x)^{-p} = \sum_{q= p}^\infty {q-1 \choose p-1} x^{q-p}.$$

Hence expanding in powers of $x$ the two sides of the identity 
$$\bigl[\prod_{i=1}^I (1-x)^{-p_i}\bigr](1-x)^{-\theta}=(1-x)^{-(\theta+P)}$$
and exchanging the order of summations in the l.h.s. we get
$$\sum_{\{q_i \geq p_i \}} \prod_{i=1}^I { q_i-1 \choose p_i-1} {\Gamma(\theta+m+P-Q) \over \Gamma(\theta)\,(m+P-Q)!} = {\Gamma(\theta+P+m) \over \Gamma(\theta+P)\,m!}.$$

The desired result is then obtained by setting $m=n-P$.

\section{Fluctuations of the sample size in the unconstrained large N limit}
\label{Fluctuations}

The multivariate generating function $\varepsilon (\{t_l \})$ was computed in Section \ref{large} after relaxing the constraint $ \sum_l l\,n_l = n$. In order to show that the constraint is automatically satisfied in the large $N$ limit we construct a generating function for the expectation value of the powers of $\nu =\sum_l l\,n_l$ in the large $N$ distribution: 
$$\eta (w) \equiv \sum_{\nu} P_{(\nu)} w^\nu = \sum_{\{N_{kl}\}} P_{\{N_{kl}\}} w^{\sum l\,n_l} = \prod_{k=1}^N  \Bigl(\sum_{\{N_{kl}\}} N_k! \prod_{l=0}^k {1 \over N_{kl}! }(P_{kl} w^l)^{N_{kl}} \Bigr),$$
and applying the multinomial formula we obtain
$$\eta (w) = \prod_{k=1}^N  \bigl(\sum_{l=0}^k P_{kl} w^l\bigr)^{N_k} =\prod_{k=1}^N (1-\rho+\rho\,w)^{kN_k} = ((1-\rho+\rho\,w)^N.$$

Expanding the result in powers of $w$ we immediately obtain $P_{(\nu)} =  P_{N\nu}$.

Since $\nu$ is distributed according to the binomial distribution, the relevant expectation values are
$$\langle {\nu \over N}\rangle  = \rho \equiv {n \over N}, \qquad \qquad \langle {\nu^2 \over N^2}\rangle -\langle {\nu \over N}\rangle ^2 = {1 \over N} \rho (1-\rho).$$

Hence fluctuations of  $\nu/N$ around $\rho$ vanish like $1/N$ in the large $N$ limit.

\section{Correlation between samples}
\label {Correlation}

An important test of randomness in sampling is offered by the measure of the correlation between two different samples. Let's consider two random samples, characterized by the sets of values $\{\hat n_a\}$ and $\{\hat m_a\}$ and by their sizes $n$ and $m$. The index $a$ labels different kinds, as in Section \ref{framework}. The correlation between the two samples is
$$C = {\sum_{a=1}^S \hat n_a \hat m_a \over \sqrt{\sum_{a=1}^S \hat n_a^2}\sqrt{\sum_{a=1}^S \hat m_a^2}}.$$

Replacing $\hat n_a$ and $\hat n_a^2$ with their expectation values, computed in Section \ref{framework},
we obtain (in the large $N$ limit)
$$\sum_{a=1}^S \langle \hat n_a\rangle  \langle \hat m_a\rangle  = n m \sum_{a=1}^S \hat p_a^2,$$
$$\sum_{a=1}^S \langle \hat n_a^2\rangle  = n^2 \bigl(\sum_{a=1}^S \hat p_a^2 +{1 \over  n}-{1 \over N}\bigr), \qquad \qquad \sum_{a=1}^S \langle \hat m_a^2\rangle  = m^2 \bigl(\sum_{a=1}^S \hat p_a^2 +{1 \over m}-{1 \over N}\bigr).$$

By making use of the results presented in Section \ref{large} we can now express the expected value of the correlation between samples in the form
$$\langle C\rangle  = {{1 \over \alpha} + {1 \over N} \over \sqrt{{1 \over \alpha} + {1 \over n}} \sqrt {{1 \over \alpha} + {1 \over m}}}.$$

For samples of equal size $n$ the expected value of the correlation takes the form $\langle C\rangle  = {n \over \alpha+ n}{\alpha +N \over N}$.
\begin{acknowledgments}

I am indebted to Sergio Caracciolo for an important observation on the r\^ole of fluctuations. I am also indebted to Roberto Dvornicich, Steve Shore and Ettore Vicari for critical reading of the manuscript.

\end{acknowledgments}

\end{document}